# A Time-driven Data Placement Strategy for a Scientific Workflow Combining Edge Computing and Cloud Computing

Bing Lin, Fangning Zhu, Jianshan Zhang, Jiaqing Chen, Xing Chen, Neal N. Xiong, Jaime Lloret Mauri

*Abstract*—Compared to traditional distributed computing environments such as grids, cloud computing provides a more cost-effective way to deploy scientific workflows. Each task of a scientific workflow requires several large datasets that are located in different datacenters from the cloud computing environment, resulting in serious data transmission delays. Edge computing reduces the data transmission delays and supports the fixed storing manner for scientific workflow private datasets, but there is a bottleneck in its storage capacity. It is a challenge to combine the advantages of both edge computing and cloud computing to rationalize the data placement of scientific workflow, and optimize the data transmission time across different datacenters. Traditional data placement strategies maintain load balancing with a given number of datacenters, which results in a large data transmission time. In this study, a self-adaptive discrete particle swarm optimization algorithm with genetic algorithm operators (GA-DPSO) was proposed to optimize the data transmission time when placing data for a scientific workflow. This approach considered the characteristics of data placement combining edge computing and cloud computing. In addition, it considered the impact factors impacting transmission delay, such as the bandwidth between datacenters, the number of edge datacenters, and the storage capacity of edge datacenters. The crossover operator and mutation operator of the genetic algorithm were adopted to avoid the premature convergence of the traditional particle swarm optimization algorithm, which enhanced the diversity of population evolution and effectively reduced the data transmission time. The experimental results show that the data placement strategy based on GA-DPSO can effectively reduce the data transmission time during workflow execution combining edge computing and cloud computing.

Xing Chen and Neal N. Xiong are both the corresponding authors.
This work was supported by the National Key R&D Program of China under Grant No. 2018YFB1004800, and the Talent Program of Fujian Province for Distinguished Young Scholars in Higher Education.

B. Lin and J. Zhang is with College of Physics and Energy, Fujian Normal University, Fujian Provincial Key Laboratory of Quantum Manipulation and New Energy Materials, Fuzhou，350117, China. The Fujian Provincial Engineering Technology Research Center of Solar Energy Conversion and Energy Storage, Fuzhou,350117, China. The Fujian Provincial Collaborative Innovation Center for Optoelectronic Semiconductors and Efficient Devices, Xia-men, 361005, China. E-mail: WheelLX@163.com, zhangjs0512@163.com
X. Chen, F. Zhu and J. Chen are with the College of Mathematics and Computer Science, Fuzhou University, Fuzhou， 350117, China. E-mail: chenxing@fzu.edu.cn, 771070906@qq.com, 330297950@qq.com.
Neal N. Xiong is with the Department of Mathematics and Computer Science, Northeastern State University, OK, USA. E-mail: xiong-naixue@gmail.com.
Jaime L. Mauri is with the Integrated Management Coastal Research Institute, Universitat Politècnica de València, Spain. E-mail: jlloret@dcom.upv.es

## I. INTRODUCTION

SCIENTIFIC applications are usually data- and computation-intensive, and they are composed of hundreds of interrelated tasks. Workflow models have been an effective way to represent complicated scientific applications, which are widely used in many scientific fields, such as astronomy [1], physics [2], and bioinformatics [3]. The complex structure and large datasets in a scientific workflow result in strict requirements on the storage capacity of the deployment environment. Grids and other traditional distributed computing environments are typically built for specific scientific research with low-level resource sharing. A scientific workflow deployed in such environments will result in more wasted resources.

Cloud computing [4,5] virtualizes resources in different geographic locations into a resource pool through virtualization technology. The resource pool is made available to end-users in a pay-as-you-go manner. Its high efficiency, flexibility, scalability, and customizable features provide a more cost-efficient way to deploy scientific workflows [6]. Cloud computing resources are usually deployed at the remote end, and the scientific workflow has large-scale datasets interaction, resulting in serious data transmission delays [7].

Edge computing resources are usually deployed in the near end, which can reduce the data transmission delays and have an effect on private datasets protection [8]. Due to the limited resources, it is impossible to store all the datasets required and generated by a scientific workflow in edge computing.

Combining the advantages of both edge computing and cloud computing to rationalize the data placement of a scientific workflow is an efficient way to reduce data transmission delays. Cloud computing ensures the resource supply and maintains the quality of service under the conditions of a drastically fluctuating workload. Edge computing guarantee the security of privacy datasets for a scientific workflow [9]. Data placement strategies for a scientific workflow combining edge computing and cloud computing have become a popular topic [10]. In the field of emergency management, a low-delay data transmission is required for a scientific workflow deployed combining edge computing and cloud computing [11]. However, the private datasets that are stored in a fixed manner lead to a large amount of data movement across datacenters during the workflow execution. There is a large contradiction between the large amount of data movement and the limited bandwidth between datacenters, resulting in serious data transmission delays. Therefore, it is important to propose a reasonable data placement strategy for a scientific workflow combining edge computing and cloud computing.

The detailed requirements of a good strategy data placement are as follows: (1) The scientific workflow has a complex structure and large datasets. Therefore, the data placement strategy should ensure high cohesion within a datacenter and low coupling between different datacenters, which reduces the data transmission time across datacenters combining edge computing and cloud computing. (2) For security reasons, private datasets should be stored in edge datacenters. Because the storage capacity of edge datacenters is limited, some datasets must be transmitted across different datacenters. It is a challenge to place the datasets with low latency un-der the conditions of the limited bandwidth and fixed private datasets.

Traditional data placement strategies for a scientific workflow mainly adopted clustering [12,13] and evolutionary algorithms [14,15]. The clustering algorithms maintained load balancing and effective resource utilization among multiple datacenters. To guarantee low-delay data transmission combining edge computing and cloud computing, a data placement strategy for a scientific workflow requires high cohesion within a datacenter and low coupling between different datacenters. However, the clustering algorithms only considered load balancing. Traditional evolutionary algorithms adopted the genetic algorithm (GA) [16], whose time complexity is very high. Therefore, a time-driven data placement strategy for a scientific workflow combining edge computing and cloud computing is still an open issue.

In previous works [17, 18], we addressed workflow scheduling based on the improved particle swarm optimization (PSO), which is an evolutionary algorithm. Workflow data placement and workflow scheduling are both NP-hard problems with many similarities. There-fore, this study proposed a self-adaptive discrete PSO algorithm with genetic algorithm operators (GA-DPSO) to reduce the data transmission time during workflow execution combining edge computing and cloud computing. This approach considered the impact factors on the transmission delay, such as the bandwidth between datacenters, the number of edge datacenters, and the storage capacity of edge datacenters.

The main contributions of this study are as follows:

1. According to the characteristics of data dependencies in a scientific workflow, preprocessing for formalizing the scientific workflow was designed to effectively compress the number of datasets and improve the execution efficiency of GA-DPSO.

2. The crossover and mutation operator of the GA were adapted to avoid the premature convergence of traditional PSO, which enhanced the diversity of population evolution and effectively reduced the data transmission time.

3. A time-driven data placement strategy based on GA-DPSO for a scientific workflow was proposed that optimized the data transmission time from a global perspective combining edge computing and cloud computing. This strategy considered the impact factors on the transmission delay, such as the bandwidth between datacenters, the number of edge datacenters, and the storage capacity of edge datacenters.

The remainder of this study is organized as follows. Related work is presented in section II. Section III discusses in detail the process of data placement for a scientific workflow combining edge computing and cloud computing, and section IV represents the proposed GA-DPSO algorithm. In section V, our algorithm is compared with other state-of-the-art algorithms. Finally, section VI summarizes the full text and presents future work.

## II. RELATED WORK

A data placement strategy for a scientific workflow is critical to the workflow system performance. Factors such as large datasets, limited bandwidth, and privacy datasets stored in fixed edge datacenters have a critical effect on data transmission time. Therefore, it is of great significance to propose a feasible data placement strategy for a scientific workflow to compress data transmission and improve system performance combining edge computing and cloud computing.

Current research mainly focused on optimizing the number of data movement and data transmission time in cloud environment. Yuan et al. [12] proposed a data placement strategy based on k-means and BEA clustering for a scientific workflow that effectively reduced the number of data movements. However, it ignored the difference in the storage capacity of each datacenter. In addition, the number of data movements did not accurately represent the amount of data movement or actual data transmission status. Wang et al. [19] designed a data placement strategy based on k-means clustering for a scientific workflow in cloud environments that considered the data size and dependency. This approach reduced the number of data movements using a data replication mechanism, but it did not formalize the data replication cost. Cui et al. [15] constructed a tripartite graph to formulate the data replica placement problem and proposed a data placement strategy based on the GA for a scientific workflow, to reduce the number and amount of data movement in cloud environments. However, this work ignored the privacy datasets in the scientific workflow. Zheng et al. [14] proposed a three-stage data placement strategy based on the GA for a scientific workflow in cloud environments that considered crucial factors such as data dependency and global load balancing across datacenters. This approach had a significant effect on the optimization of the data transmission time. However, it had high time complexity. Li et al. [13] proposed a data placement strategy based on data dependency destruction for a scientific workflow in hybrid cloud environments that effectively reduced data transmission time across different datacenters. This work has influenced the present study, yet it ignored the difference in storage capacity across different datacenters and the different bandwidths between datacenters.

Edge computing has recently emerged as an important paradigm to bring computation and cache resources to the edge of core networks [16]. Recently, there were many studies aiming at improving QoS in edge computing. Gang Sun et al. designed DMRT_SL and DMRT_NSL algorithms to efficiently reduce the latency for provisioning the workflow in edge computing-like service request, which met the requirements of different kinds of service requests [20]. This strategy ignored the impact of different bandwidths across multiple datacenters on data placement. [21] presented a workflow-net-based mechanism for mobile edge node cooperation in fog-cloud networks to form guaranteed service specific overlays for faster service delivery. By proposing an algorithm that predicts the response time of complex event processing (CEP) services dynamically, [22] deployed the operators on the edge nodes with the minimum predicted delay to reduce the response time. It ignored the impact of different datacenter storage capacities on the data placement. While some research focused on reducing energy consumption. [23] designed an energy-efficient computation offloading (EECO) scheme, which jointly optimized offloading and radio resource allocation to obtain the minimal energy consumption under the latency constraints.

Combining edge computing and cloud computing can solve delay minimization problem effectively. Odessa [24] was an example that could offload tasks to either the cloud or a dedicated edge computing cloudlet. Odessa could adapt quickly to changes in scene complexity, compute resource availability, and network bandwidth. But it did not make good use of the public cloud. Both [16] and [25] considered the characteristics of data placement combining edge computing and cloud computing. The former research mainly focused on putting forward a heuristic algorithm based on genetic algorithm (GA) and simulated annealing (SA) to solve a resource-constrained delay minimization problem, the latter one focused on proposing a cloud assisted mobile edge computing (CAME) framework to solve a capacity-constrained delay minimization problem. [26] introduced strategies to create placement configurations for data stream processing applications whose operator topologies follow series parallel graphs, aiming at improving the response time. The similarity between their work and ours is that both the placement decisions took cloud computing and edge computing into consideration, but their work focused on data stream processing.

In summary, previous studies have researched the data placement for a scientific workflow. However, they mostly ignored crucial factors such as the limited storage capacity of edge cloud datacenters and the difference in bandwidths across different datacenters on the data placement combining edge computing and cloud computing.

III. PROBLEM DEFINITION AND ANALYSIS

The core purpose of data placement for a scientific workflow is to achieve a minimum data transmission time while satisfying the storage capacity constraint of each datacenter. In this section, we define the concepts related to the data placement strategies for a scientific workflow combining edge computing and cloud computing and analyze the data transmission time optimization using a specific example.

*A. Problem Definition*

The problem definition includes a new hybrid environment combining edge computing and cloud computing, a scientific workflow, and a data placement strategy.

The hybrid environment combining edge computing and cloud computing $DC = \{DC_{cld}, DC_{edg}\}$ includes cloud computing at the remote end and edge computing in the near end, which both consist of multiple datacenters. Cloud computing $DC_{cld} = \{dc_1, dc_2, ..., dc_n\}$ consists of n datacenters, and edge computing $DC_{edg} = \{dc_1, dc_2, ..., dc_m\}$ consists of m datacenters. This study designs a data placement strategy. Thus, we focus on the storage capacity of each datacenter and ignore their computing capacity. The datacenter $dc_i$ (whose number is $i$) is expressed as

$$dc_i = <capacity_i, type_i>, \quad (1)$$

where $capacity_i$ represents the storage capacity of the datacenter $dc_i$, and the datasets stored in this datacenter cannot exceed its capacity. $type_i = \{0, 1\}$ represents the location that the datacenter $dc_i$ belongs to. When $type_i = 0$, $dc_i$ belongs to cloud computing, and it can only store public datasets. When $type_i = 1$, $dc_i$ belongs to edge computing, and it can store both private and public datasets. The bandwidth across different datacenters is expressed as follows.

$$Bandwidth = \begin{bmatrix} b_{11} & b_{12} & \cdots & b_{1|DC|} \\ b_{21} & b_{22} & \cdots & b_{2|DC|} \\ \vdots & \vdots & \cdots & \vdots \\ b_{|DC|1} & b_{|DC|2} & \cdots & b_{|DC||DC|} \end{bmatrix}, \quad (2)$$

$$b_{ij} = <band_{ij}, type_i, type_j>, \quad (3)$$

where $b_{ij}$ represents the bandwidth between datacenters $dc_i$ and $dc_j$. $band_{ij}$ is the measured value of bandwidth $b_{ij}$, where $\forall i, j = 1, 2, ..., |DC|$ and $i \neq j$. The bandwidth is assumed to be known and not fluctuate.

The scientific workflow is represented by a directed acyclic graph $G = (T, E, DS)$ [21], where $T = \{t_1, t_2, ..., t_r\}$ denotes a set of nodes containing $r$ tasks, $E = \{e_{12}, e_{13}, ..., e_{ij}\}$ denotes the data dependencies between each pair of tasks, and $DS = \{ds_1, ds_2, ..., ds_n\}$ denotes all datasets in the scientific workflow.

Each data-dependent edge $e_{ij} = (t_i, t_j)$ represents a data dependency between task $t_i$ and task $t_j$, where task $t_i$ is the direct precursor of task $t_j$, and task $t_j$ is the direct successor of task $t_i$. In the process of scheduling a scientific workflow, a task cannot start until all of its precursors have been completed.

For a task $t_i = <IDS_i, ODS_i>$, $IDS_i$ is the input datasets of $t_i$, and $ODS_i$ is the output datasets of $t_i$. The relationship between the task set and dataset is many-to-many (that is, one data may be used by multiple tasks, and one task may also require multiple input datasets).

For a dataset $ds_i = <dsize_i, gt_i, lc_i, flc_i>$, $dsize_i$ represents the dataset size, $gt_i$ represents the task generating $ds_i$ using (4), $lc_i$ represents the original storage location of $ds_i$ using (5), and $flc_i$ represents the final placement location of $ds_i$.

$$gt_i = \begin{cases} 0, & ds_i \in DS_{ini} \\ Task(ds_i), & ds_i \in DS_{gen} \end{cases}, \quad (4)$$

$$lc_i = \begin{cases} 0, & ds_i \in DS_{flex} \\ fix(ds_i), & ds_i \in DS_{fix} \end{cases}. \quad (5)$$

Datasets can be divided into initial datasets $DS_{ini}$ and generated datasets $DS_{gen}$ according to data sources. The initial datasets are the input datasets of a scientific workflow, and the generated datasets are the intermediate datasets generated during the scientific workflow execution. In (4), $Task(ds_i)$ represents the task generating the dataset $ds_i$. In addition, datasets can also be divided into fixed datasets $DS_{fix}$ (that is, private datasets) and flexible datasets $DS_{flex}$ (that is, public datasets) according to their storage locations. Private datasets can only be stored in edge datacenters, and $fix(ds_i)$ represents the edge datacenter storing the private dataset $ds_i$.

The purpose of our data placement strategy is to minimize the data transmission time while satisfying all requirements during workflow execution. Any task execution in a workflow must satisfy two conditions: (1) The task should be scheduled to a specific datacenter. (2) The input datasets required by the task are already in the specific datacenter. Because the time for scheduling tasks to datacenters is much less than the time for transmitting datasets from one datacenter to another [27, 28], this study only focuses on the data transmission time. Assume that a task is scheduled to the datacenter with a minimum data transmission time after data placement for a scientific workflow. The data placement can be defined as $S = (DS, DC, Map, T_{total})$, where $Map = \bigcup_{i=1,2,...,|DS|} \{<dc_i, ds_k, dc_j>\}$ represents the maps from the datasets $DS$ to the datacenters $DC$. A map $<dc_i, ds_k, dc_j>$ represents the dataset $ds_k$ transmission from the original storage location $dc_i$ to the final placement location $dc_j$, and the data transmission time is calculated as (6). $T_{total}$ represents the total data transmission time during data placement for a scientific workflow, which is shown in (7).

$$T_{transfer}(dc_i, ds_k, dc_j) = \frac{dsize_k}{band_{ij}}, \quad (6)$$

$$T_{total} = \sum_{i=1}^{|DC|} \sum_{j \neq i}^{|DC|} \sum_{k=1}^{|DS|} T_{transfer}(dc_i, ds_k, dc_j) \cdot e_{ijk}, \quad (7)$$

where $e_{ijk} = \{0, 1\}$ represents if there is a dataset $ds_k$ transmitted from the original storage location $dc_i$ to the final placement location $dc_j$, $e_{ijk} = 1$ indicates presence, and $e_{ijk} = 0$ indicates absence.

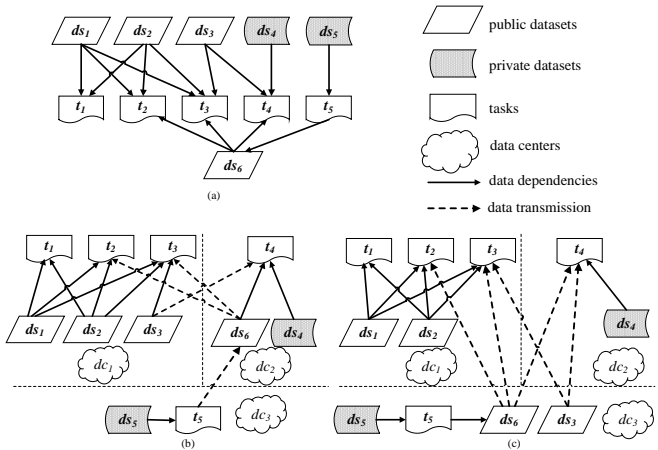

Fig. 1. A sample of data placement for a scientific workflow.

The problem of time-driven data placement strategies for a scientific workflow combining edge computing and cloud computing can be formalized as (8). Its core purpose is to pursue a minimum total data transmission time while satisfying the storage capacity constraint for each datacenter.

$$\begin{aligned}&\textbf{Minimize } T_{total}\\&\textbf{subject to } \forall i, \sum_{j=1}^{|DS|} ds_j \cdot u_{ij} \leq capacity_i,\end{aligned} \quad (8)$$

where $u_{ij} = \{0, 1\}$ indicates whether the dataset $ds_j$ is stored in datacenter $dc_i$. $u_{ij} = 1$ if yes and $u_{ij} = 0$ if no.

### B. Problem Analysis

Figure 1(a) is a sample of data placement for a scientific workflow, which includes five tasks $\{t_1, t_2, t_3, t_4, t_5\}$, five input datasets $\{ds_1, ds_2, ds_3, ds_4, ds_5\}$, and an intermediate dataset $\{ds_6\}$. These dataset sizes $\{dsize_1, dsize_2, dsize_3, dsize_4, dsize_5, dsize_6\}$ are $\{3GB, 5GB, 3GB, 3GB, 5GB, 8GB\}$, respectively, and $ds_4$ is the private dataset that is only stored in edge datacenter $dc_2$. The input datasets of task $t_4$ are $\{ds_3, ds_4, ds_6\}$, which include $ds_4$. Therefore, task $t_4$ must be executed in datacenter $dc_2$. Similarly, dataset $ds_5$ is private and only stored in edge datacenter $dc_3$. Task $t_5$ must be executed in datacenter $dc_3$. Two data placement results with different strategies are shown in Figures 1(b) and 1(c), where $dc_1$ is a cloud datacenter with unlimited storage capacity, and the other two datacenters ($dc_2$ and $dc_3$) are edge datacenters with the same storage capacity (20 GB). The bandwidth between edge datacenters is approximately 10 times faster than the bandwidth between a cloud datacenter and an edge datacenter [29]. Assume that the bandwidth $\{band_{12}, band_{13}, band_{23}\}$ across three datacenters is $\{10 M/s, 20 M/s, 150 M/s\}$.

Figure 1(b) is the data placement result according to [13]. Based on the partitioning model of the dependency matrix, the public datasets $\{ds_1, ds_2, ds_3\}$ are stored in cloud datacenter $dc_1$, and $ds_6$ is stored in edge datacenter $dc_2$. The privacy datasets $\{ds_4, ds_5\}$ are stored in their corresponding edge datacenters. This data placement result is that the number of data movements is 4, the amount of data movement is 27 GB, and the data transmission time is approximately 1953 s.

Figure 1(c) is the optimal data placement result. The public datasets $\{ds_1, ds_2\}$ are stored in cloud datacenter $dc_1$, and the datasets $\{ds_3, ds_6\}$ are stored in edge datacenter $dc_3$. The data placement result is that the number of data movements is 5, the amount of data movement is 30 GB, and the data transmission time is approximately 1023 s. Due to the consideration of the bandwidth across different datacenters, the data transmission time of this strategy is significantly better than the former in [13].

The traditional matrix-partitioning model [12] tends to place datasets with high data dependency in the same datacenter, which effectively reduces the amount of data movement across different datacenters. However, these approaches ignore the impact of bandwidth on the final data placement when pursuing a short data transmission time. This study proposed a data placement strategy based on GA-DPSO, which adaptively placed datasets while considering the bandwidth between datacenters, number of edge datacenters, and storage capacity of edge datacenters.

## IV. DATA PLACEMENT STRATEGY BASED ON GA-DPSO

For a data placement strategy $S = (DS, DC, Map, T_{total})$, its core purpose is to find the best map from $DS$ to $DC$ that has a minimum data transmission time $T_{total}$. It is an NP-hard problem to find the best map from $DS$ to $DC$ [30]. Therefore, we proposed a data placement strategy based on the GA-DPSO algorithm to optimize the data transmission time from a global perspective combining edge computing and cloud computing. To improve the strategy efficiency, a preprocessing of compressing datasets was performed. The preprocessing and GA-DPSO algorithm are described as follows.

### A. Preprocessing for a scientific workflow

**Algorithm 1**: Merge each cut-dataset into a new dataset

**procedure** preprocess ($G(T, E, DS)$)
   1: Record the out-degree and in-degree of $G$'s datasets
   2: Find all cut-edge datasets.
   3: If there are cut-edge datasets, merge each cut-edge dataset into a new dataset.
   4: Repeat *step* **2** until there is no cut-edge dataset.
**end procedure**

Algorithm 1 introduces the preprocessing pseudocode for a scientific workflow that merges each cut-edge dataset into a new one. A cut-edge dataset is one where there are two adjacent datasets (such as $ds_i$ and $ds_j$), at least one dataset is public, and they only have one common task. The out-degree of $ds_i$ is 1 and the in-degree is 1, and there is only one task between $ds_i$ and $ds_j$. The process of merging a cut-edge dataset into a new one is shown in Figure 2(a). The science workflow Epigenomics [31] have many cut-edge datasets, and the number of datasets is compressed by more than 30% after preprocessing. Figure 2(b) shows the structure of the Epigenomics before and after preprocessing. GA-DPSO will process a workflow faster with less datasets.

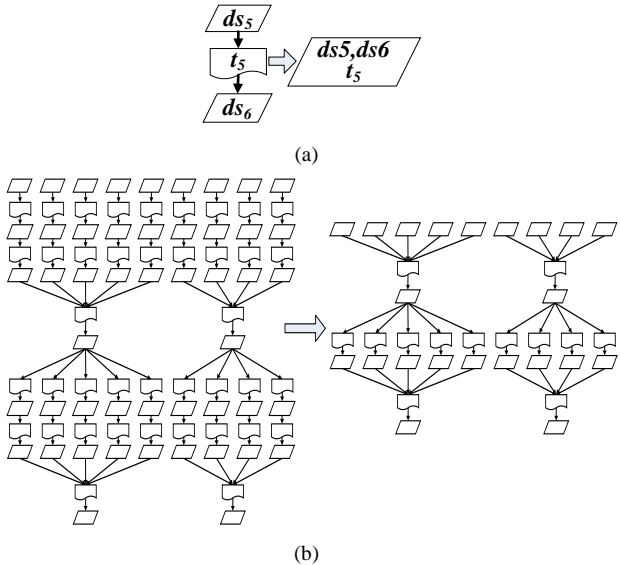

Fig. 2 Preprocessing for a scientific workflow: (a) Merging a cut-edge dataset into a new one; (b) The structure of Epigenomics before and after preprocessing

**Property 1**: Preprocessing compresses the number of datasets in a scientific workflow and improves the execution efficiency of GA-DPSO. However, it may affect the final data placement result.

The number of datasets is compressed as shown in Figure 2. Part B in this section introduces the problem encoding of GA-DPSO, whose dimensions are based on the number of datasets. Therefore, compressing the number of datasets reduces the coding dimension of GA-DPSO, which will improve the execution efficiency. In Figure 2(a), $ds_5$ and $ds_6$ are merged together. This means that $ds_5$ and $ds_6$ must be stored in the same datacenter after preprocessing. Without preprocessing, $ds_5$ and $ds_6$ may be stored in different edge datacenters. Therefore, the preprocessing may affect the final result of data placement.

### B. GA-DOSO

PSO is an evolutionary computation technique inspired by the social behavior of bird flocks, which was first presented by Kennedy and Eberhart [32]. The particle is the most important concept in PSO. A particle represents a candidate solution that moves around in the search-space. Each particle has its own velocity, which determines its future direction and magnitude. The movement of each particle is determined by its velocity and position, and they iteratively update these using (9) and (10).

$$V_i^{t+1} = w \times V_i^t + c_1 r_1(pBest_i^t - X_i^t) + c_2 r_2(gBest^t - X_i^t), \quad (9)$$

$$X_i^{t+1} = X_i^t + V_i^{t+1}. \quad (10)$$

$V_i^t$ and $X_i^t$ represent the velocity and position of the $i^{th}$ particle at the $t^{th}$ iteration, respectively. In general, a maximum velocity $V_{max}$ is defined to ensure that the particle search-space is in the range of the solution space. This velocity is affected by the personal best position of the particle, $pBest$, and the global best position of the population, $gBest$. The inertia weight $w$ determines how much the previous velocity can affect the current velocity. It has a significant impact on the convergence of the algorithm. The two acceleration coefficients (that is, $c_1$ and $c_2$) represent the particle cognitive ability to their personal and global best values. To enhance the randomness of searching, the algorithm introduces two random numbers ($r_1$ and $r_2$) whose values are both between 0 and 1. In addition, a fitness function is used to evaluate the quality of a particle.

Traditional PSO is used to solve the continuous problem. The data placement problem in this study is discrete and requires a new problem-coding approach. For the premature convergence of traditional PSO, a new update strategy for particles is needed. In addition, the parameter setting may affect the search capability of an evolutionary algorithm. Therefore, GA-DPSO is proposed to solve the above problems. The data placement strategy based on GA-DPSO is described in detail as follows.

**1) Problem encoding**

To improve the algorithm performance and enhance its searching efficiency, a good encoding strategy should satisfy the following three principles [33]:

**Definition** 1 (**Completeness**). Each candidate solution in the problem space can be encoded as a particle.

**Definition** 2 (**Non-redundancy**). A candidate solution in the problem space has only one corresponding encoded particle.

**Definition 3** (**Viability**). Each encoded particle corresponds to a candidate solution in the problem space.

It is difficult to propose an encoding strategy that satisfies the above three principles. Inspired by [34], we adopt the discrete encoding strategy to generate n-dimensional candidate solution particles. A particle represents a data placement solution for a scientific workflow combining edge computing and cloud computing, and the $i^{th}$ particle in the $t^{th}$ iteration is shown in (11).

$$X_i^t = (x_{i1}^t, x_{i2}^t, \ldots, x_{in}^t), \quad (11)$$

where n is the number of datasets after preprocessing, and each particle is an integer-valued vector of dimension n. $x_{ik}$ ($k=1, 2, \ldots, n$) represents the final placement location of the kth dataset in the $tth$ iteration, whose value is the datacenter number, that is, $x_{ik} = \{1, 2, ..., |DC|\}$. Note that the storage location of the private datasets is fixed, which is never changed. For example, in Figure 1, $ds_4$ and $ds_5$ can only be fixed and stored in $dc_2$ and $dc_3$, respectively. Figure 3 shows an encoded particle corresponding to the data placement of Figure 1(c). After preprocessing, the number of datasets is changed from six to five. The datasets $ds_5$ and $ds_6$ are compressed into a single dataset stored in $dc_3$.

Fig. 3 An encoded particle corresponding to the data placement

**Property 2**: Our discrete encoding strategy satisfies the non-redundancy and completeness principles, but does not satisfy the viability principle.

After data placement, each dataset is stored in the corresponding datacenter, which has a corresponding datacenter number. The final placement location of a dataset can only be in a datacenter. A data placement strategy for a scientific workflow corresponds to an n-dimensional particle. The value of the $i^{th}$ dimension in a particle is the datacenter number that stores the $i^{th}$ dataset. A data placement strategy only corresponds to one encoded particle, which satisfies the non-redundancy principle. Each public dataset can be stored in different datacenters, and the value of corresponding dimensions in a particle can be a different datacenter number. Each data placement strategy has the corresponding encoded particle, which satisfies the completeness principle. Some encoded particles cannot be the candidate solutions for the problem space. If the final placement location of datasets in Figure 3 is (1, 2, 2, 2, 2), then all datasets except $ds_1$ are stored in $dc_2$. The size of datasets in $dc_2$ is 24 GB, which exceeds its storage capacity (that is, 20 GB). Therefore, the discrete encoding strategy does not satisfy the viability principle.

**2) Fitness function**

A fitness function evaluates the advantages and disadvantages of a particle. In general, a particle with smaller fitness has better performance [35]. The purpose of this study is to reduce the transmission time of data placement for a scientific workflow. The smaller the data transmission time, the better the particle. The fitness function is equal to the transmission time of a data placement strategy corresponding to a specific particle. However, our discrete encoding strategy does not satisfy the viability principle, and the fitness function must be defined according to different situations.

**Definition 4** (**Feasible particle**). An encoded particle (which corresponds to a specific data placement strategy) satisfies the storage capacity constraint. That is, there is no edge datacenter exceeding its storage capacity.

**Definition 5** (**Infeasible particle**). An encoded particle (which corresponds to a specific data placement strategy) does not satisfy the storage capacity constraint. That is, there is at least one edge datacenter exceeding its storage capacity.

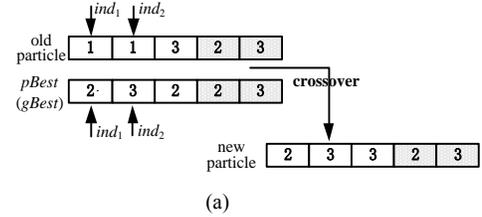

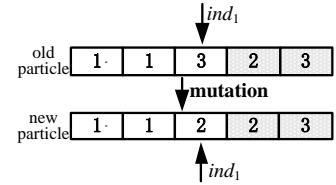

Fig. 4 Update operation: (a) Crossover operator for the individual (social) cognition component; (b) Mutation operator for the inertia component

We compare the value of the fitness function of two encoded particles for three different cases.

**Case 1**: Both encoded particles are feasible, and the particle with the smaller data transmission time is selected as the better one. The fitness function is defined as follows.

$$fitness = T_{total(X_i)}. \quad (12)$$

**Case 2**: Both encoded particles are infeasible, and the particle with the smaller data transmission time is selected as the better one. An infeasible particle may become a feasible particle after the update operation, and the particle with the smaller data transmission time is more likely to be selected. Therefore, the fitness function is consistent with (12).

**Case 3**: An encoded particle is infeasible, and another one is feasible. There is no doubt that the feasible particle is selected, and the fitness function is defined as follows.

$$fitness = \begin{cases} 0, \text{ if } \forall i, \sum_{j=1}^{|DS|} ds_j \cdot u_{ij} \leq capacity_i \\ 1, \text{else} \end{cases}. \quad (13)$$

**3) Update strategy**

As shown in (9), traditional PSO includes three main parts: *inertia*, *individual cognition*, and *social cognition*. The movement of each particle is influenced by its personal best-known position, but is also guided toward the global best-known position in the search-space [36]. The traditional PSO is easy to prematurely converge into a local optimum. To enhance the search ability of our strategy, we adapt the crossover and mutation operators of the GA for particle update to explore a wider range of the solution space. The update strategy for the $i^{th}$ particle at the $t^{th}$ iteration is described as follows.

$$X_i^t = c_2 \oplus C_g(c_1 \oplus C_p(w \oplus M_u(X_i^{t-1}), pBest_i^{t-1}), gBest^{t-1}), \quad (14)$$

where $C_g()$, $C_p()$ are both crossover operators, and $M_u()$ represents the mutation operator.

For the individual cognition and social cognition components, we adapt the crossover operator of the GA and update the corresponding parts of (9), which is shown in (15) and (16).

$$B_i^t = c_1 \oplus C_p(A_i^t, pBest^{t-1}) = \begin{cases} C_p(A_i^t, pBest^{t-1}) & r_1 < c_1 \\ A_i^t & else \end{cases}, \quad (15)$$

$$C_i^t = c_2 \oplus C_g(B_i^t, gBest^{t-1}) = \begin{cases} C_g(B_i^t, gBest^{t-1}) & r_2 < c_2 \\ B_i^t & else \end{cases}, \quad (16)$$

where $r_1$ (or $r_2$) is a random factor between 0 and 1. $C_p()$ (or $C_g()$) randomly selects two indexes in an old particle, and replaces the segment between them with the one in the *pBest* (or *gBest*) particle. Figure 4(a) illustrates the crossover operator for the *individual* (or *social*) cognition component. It randomly selects the two crossover indexes ($ind_1$ and $ind_2$), and replaces the segment between 1st ($ind_1$) index and 2nd ($ind_2$) index in the old particle with the *pBest* (or *gBest*) particle.

**Property 3**: The crossover operator may change an encoded particle from feasible to infeasible, and vice versa.

The encoded particle (1, 1, 3, 2, 3) in Figure 3 is feasible. Assume that the *pBest* particle is (2, 3, 2, 2, 3), and the crossover indexes are 1st and 2nd. Therefore, the generated encoded particle is (2, 3, 3, 2, 3) after the crossover operator. This particle places $\{ds_2, ds_3, ds_5, ds_6\}$ in $dc_3$, and the size of all datasets in $dc_3$ is 21 GB, which exceeds the storage capacity of $dc_3$ (20 GB). This generated particle is infeasible. On the contrary, an infeasible particle (2, 3, 3, 2, 3) crossover with the *pBest* particle (2, 2, 1, 2, 3) in index 1st and 2nd. The new generated particle (2, 2, 3, 2, 3) is feasible.

For the inertia component, we adapt the mutation operator of the GA and update the inertia part of (9), which is shown in (17).

$$A_i^t = w \oplus M_u(X_i^{t-1}) = \begin{cases} M_u(X_i^{t-1}) & r_3 < w \\ X_i^{t-1} & else \end{cases}, \quad (17)$$

where $r_3$ is a random factor between zero and one. Because the private datasets are stored in the corresponding fixed datacenters, $M_u()$ randomly selects an index in an old particle, which can only be within the position of public datasets. $M_u()$ then randomly changes this index value in the range of the datacenter number. The mutation operator selects the index in two cases.

**Case 1**: The old particle is feasible. $M_u()$ randomly changes this index value in the range of the datacenter number.

**Case 2**: The old particle is infeasible. $M_u()$ randomly selects one index of the overloaded datacenters, and then randomly changes this index value in the range of the datacenter number.

The encoded particle in Figure 3 belongs to **Case 1**. $M_u()$ randomly selects the index $ind_1$, and then updates the value of $ind_1$ from 3 to 2 in Figure 4(b).

**Property 4**: The mutation operator may change an encoded particle from feasible to infeasible, and vice versa.

The mutation operator randomly selects 2nd index of a feasible particle (1, 2, 3, 2, 3) to mutate, and then generates a new infeasible particle (1, 3, 3, 2, 3). This new particle stores $\{ds_2, ds_3, ds_5, ds_6\}$ in $dc_3$, whose size of datasets is 21 GB, exceeding its storage capacity (20 GB). Alternately, it mutates an infeasible particle (1, 3, 3, 2, 3) in index 2nd, and then generates a new feasible particle (1, 1, 3, 2, 3).

### 4) A map from a particle to a data placement

**Algorithm 2**: A map from a particle to a data placement

```
procedure dataPlacement (G, DC, X)
1: Initialization: dc_cur(i) ← 0, T_total ← 0.
2: foreach ds_i of DS_ini // Determine whether there is an overloaded
datacenter during placing initial datasets
3:     dc_cur(X[i]) += dsize_i, place ds_i in dc_X[i]
4:     if dc_cur(X[i]) > capacity_X[i] then
5:         return this particle is infeasible
6:     end if
7: end for
8: for j = 1 to j = |T| // Determine whether there is an overloaded datacenter
during tasks execution
9:     Place task t_j in datacenter dc_j with minimal data transmission time
10:    if dc_cur(j)+sum(IDS_j)+sum(ODS_j) > capacity_j
11:        return this particle is infeasible
12:    end if
13:    Place the output datasets ODS_j of t_j in the corresponding datacenters,
and update their current storage
14: end for
15: for j = 1 to j = |T| // Calculate the total transmission time of data
placement
16:    Find the datacenters DC_j storing the input datasets IDS_j of t_j
17:    Calculate the transmission time from IDS_j to dc_j according to (6)
18:    T_total += Transfer_j
19: end for
20: Output the data placement strategy and the corresponding T_total
end procedure
```

Algorithm 2 is the pseudocode of mapping a particle to a data placement for a scientific workflow with inputs, including a scientific workflow $G = (T, E, DS)$, the datacenters $DC$, and the encoded particle $X$. First, the current storage of all datacenters $dc_{cur(i)}$ is set to 0 and the total data transmission time $T_{total}$ is set to 0 (line 1). After initialization, the datasets are stored in the corresponding datacenters, and the current storage of each datacenter $dc_{cur(X[i])}$ is recorded. If the storage of any edge datacenter exceeds its storage capacity, then the encoded particle is infeasible and returned (line 2-7). According to the task execution sequence, the task $t_j$ is placed in datacenter $dc_j$ with a minimal transmission time. If the sum (including the current storage of $dc_j$), the size of input datasets of task $t_j$, and the size of output datasets of task $t_j$ exceeds the storage capacity of $dc_j$, then the encoded particle is infeasible and returned. Otherwise, the output datasets of $t_j$ are stored in the corresponding datacenters, whose current storage is updated (lines 8-14). If the encoded particle is feasible, we further calculate the data transmission time. All tasks are sequentially scanned, and the datacenters $DC_j$ that store the input datasets $IDS_j$ of tj are identified. The transmission time $Transfer_j$ from $IDS_j$ to $dc_j$ according to (6) is calculated, and all related transmission times are superimposed to calculate the total data transmission time $T_{total}$ (lines 15-19). Finally, the data placement strategy and corresponding $T_{total}$ are output (line 20).

### 5) Parameter settings

The inertia weight $w$ in (9) determines the speed change, which has an effect on the search ability and convergence of PSO [37]. When the inertia weight w is large, the global search ability of PSO is strong and does not easily converge; otherwise, the local search ability of PSO is strong and converges easily. Equation (18) is a classical adjustment mechanism of the inertia weight [38]. In the initial stage of PSO, more focus is placed on the global search to a wider range of solution spaces. As the number of subsequent iterations increases and the search goes deeper, PSO focuses more on the local search ability. Therefore, the value of inertia weight $w$ decreases linearly with the number of iterations, where $w_{max}$ and $w_{min}$ are the maximum and minimum values of $w$, respectively, during the initialization phase. $iters_{max}$ and $iters_{cur}$ are the maximum and the current number of iterations, respectively.

$$w = w_{max} - iters_{cur} \times \frac{w_{max} - w_{min}}{iters_{max}}. \quad (18)$$

The inertia weight of (18) is adjusted based on the number of iterations, which does not satisfy the nonlinear characteristics of data placement. Therefore, an inertia weight that can adaptively adjust the search ability according to the current particle quality is designed in (19). The new adjustment mechanism can adaptively adjust its search ability according to the difference between the current and global best particles.

$$w = w_{max} - (w_{max} - w_{min}) \times \exp(d(X_i^{t-1})/(d(X_i^{t-1})-1.01)), \quad (19)$$

$$d(X^{t-1}) = \frac{div(X^{t-1}, gBest^{t-1})}{|DS|}, \quad (20)$$

where $div(X^{t-1}, gBest^{t-1})$ represents the number of different values between the current particle $X^{t-1}$ and the global best particle $gBest^{t-1}$. When $div(X^{t-1}, gBest^{t-1})$ is large (which means that there is a big difference between $X^{t-1}$ and $gBest^{t-1}$), then it must enhance the global search ability. Therefore, the weight of $w$ should be increased to ensure a larger search range and avoid premature convergence. Otherwise, it must enhance the local search ability and accelerate the convergence to find an optimal solution.

According to the linear increase (or decrease) strategy [39], the other two acceleration coefficients ($c_1$ and $c_2$) are defined as (21) and (22).

$$c_1 = c_1^{start} - \frac{c_1^{start} - c_1^{end}}{iters_{max}} \times iters_{cur}, \quad (21)$$

$$c_2 = c_2^{start} - \frac{c_2^{start} - c_2^{end}}{iters_{max}} \times iters_{cur}. \quad (22)$$

Note that $c_1^{start}$ and $c_1^{end}$ are the initial and final values of $c_1$. $c_2^{start}$ and $c_2^{end}$ are the initial and final values of $c_2$.

6) **Algorithm flowchart**

Figure 5 is the GA-DPSO flowchart, whose detailed steps are described as follows.

***Step 1***: Compress the number of datasets according to the preprocessing for a science workflow in part 1 in this section (that is, Algorithm 1).

***Step 2***: Initialize relevant parameters of GA-DPSO such as population size, maximum iteration, inertia weight, and cognitive factors, then randomly generate the initial population.

***Step 3***: According to the map from a particle to a data placement in part 2 in this section (that is, Algorithm 2), calculate the fitness of each particle based on (12) and (13). Each particle is set as its personal best particle, and the particle with the smallest fitness is set as the global best particle of the population.

***Step 4***: Update particles based on (14) - (17), and recalculate the fitness of each updated particle.

***Step 5***: If the fitness of the updated particle is smaller than its personal best particle, then set the updated particle as its own

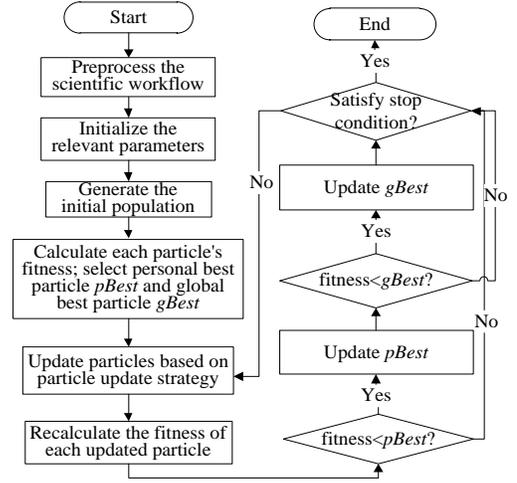

Fig. 5 GA-DPSO flowchart

personal best particle. Otherwise, go to ***Step 7***.

***Step 6***: If the fitness of the updated particle is smaller than the global best particle, then set the updated particle as the global best particle.

***Step 7***: Verify whether the stop condition is met. If it is not satisfied, then go to ***Step 4***. Otherwise, terminate the procedure.

V. EXPERIMENTAL RESULTS AND ANALYSIS

We conducted all simulation experiments on a Win8 64-bit operating system with an i7-7500U 2.90 GHz Intel (R) Core (TM) processor and 8GB of RAM. According to [38], the relevant parameters of GA-DPSO were set as follows. The size of initial population was 100, the maximum iteration was 1000, $w_{max} = 0.9$, $w_{min} = 0.4$, $c_1^{start} = 0.9$, $c_1^{end} = 0.2$, $c_2^{start} = 0.9$, and $c_2^{end} = 0.4$.

*A. Experimental setup*

We conducted our experiments using five types of partly synthetic workflows: CyberShake in earthquake science, Montage in astronomy, SIPHT in bioinformatics, Epigenomics in biogenetics, and LIGO in gravitational physics. These were all investigated in depth by Bharathi et al. [30]. Both the number of datasets and the structure in each type of scientific workflow are different. The detailed information about dependency structure and input/output datasets for each type of workflows is recorded in an XML file[1]. For each scientific field, there are four kinds of scientific workflows with different sizes of tasks, from which this study selected three for our experiments: small (approximately 30 tasks), medium (approximately 50 tasks), and large (approximately 1000 tasks).

We evaluate the effect of several impact factors on different data placement strategies. Therefore, we adjust some impact factors based on the basic experiment, whose setup is described as follows. The hybrid environment consists of four datacenters $\{dc_1, dc_2, dc_3, dc_4\}$, where $dc_1$ is a cloud datacenter with unlimited storage capacity, and the other three datacenters are edge datacenters. We define the benchmark storage capacity $cap_{benchmark}$ as (23), and the storage capacity of three edge datacenters is 2.6 times that of $cap_{benchmark}$.

---
[1] https://confluence.pegasus.isi.edu/display/pegasus/WorkflowGenerator

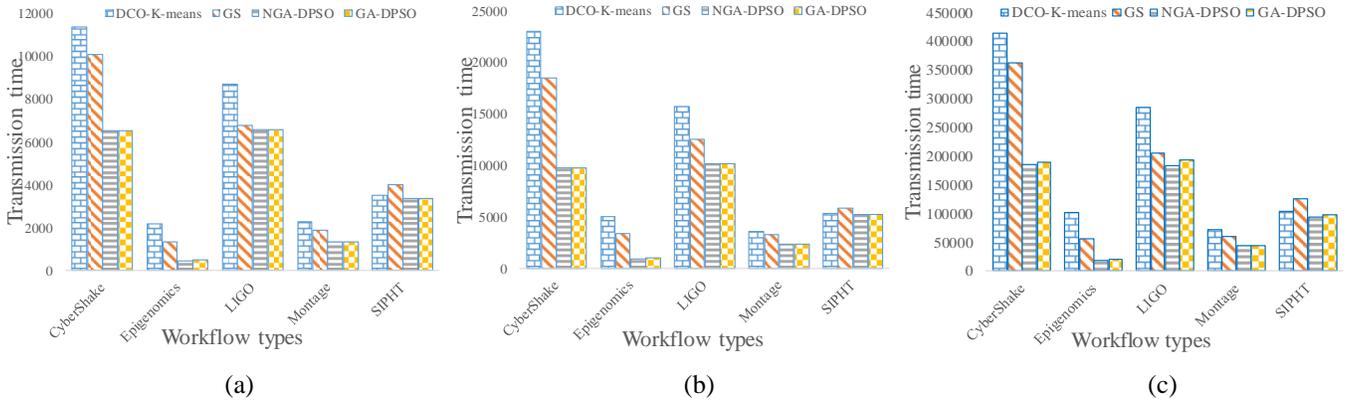

Fig. 6 Data transmission time of different strategies for three kinds of workflows in basic experiment: (a) Small; (b) Medium; (c) Large

$$capacity = \frac{\sum_{i=1}^{|DS|} dsize_i}{|DC|-1}. \quad (23)$$

The proportion of private datasets in a workflow is set to 25%, and the bandwidth across different datacenters is described as follows (its unit is M/s).

$$Bandwidth = \begin{bmatrix} \sim & 10 & 20 & 30 \\ 10 & \sim & 150 & 150 \\ 20 & 150 & \sim & 100 \\ 30 & 150 & 100 & \sim \end{bmatrix}. \quad (24)$$

### B. Competitive algorithms

There are certain similarities between the hybrid cloud environment and the environment combining edge computing and cloud computing [29]. To verify the effectiveness of GA-DPSO, we modified the DCO-k-means data placement strategy [13] and the GA-based data placement strategy (GS) [15] to adapt the time-driven data placement strategies for a scientific workflow combining edge computing and cloud computing.

The DCO-k-means data placement strategy first clustered the datasets according to the data dependency, and then divided the datasets into data blocks using a matrix-partitioning model. The data dependency degree, which represented the number of tasks that simultaneously accepted two relevant datasets as input, played a significant role in the matrix-partitioning model. The definition of data dependency degree ignored the factor of bandwidth while optimizing data transmission time. Therefore, we redefine the data dependency degree $dependency_{ij}$ as follows.

$$dependency_{ij} = Count(ds_i.T \cap ds_j.T) \cdot$$

$$\begin{cases} \dfrac{\min(dsize_i, dsize_j)}{band_{(flc_i)(flc_j)}}, & ds_i, ds_j \in DS_{flex} \\ \dfrac{dsize_i}{band_{(flc_i)(flc_j)}}, & ds_i \in DS_{flex}, ds_j \in DS_{fix} \\ \dfrac{dsize_j}{band_{(flc_i)(flc_j)}}, & ds_i \in DS_{fix}, ds_j \in DS_{flex} \\ 0, & ds_i, ds_j \in DS_{fix} \end{cases} \quad (25)$$

where $Count(ds_i.T \cap ds_j.T)$ represents the number of tasks that accept both datasets $ds_i$ and $ds_j$ as input, and $band_{(flc_i)(flc_j)}$ represents the pre-placement bandwidth between the datacenter storing $ds_i$ and the one storing $ds_j$. The new definition of data dependency degree considers the influence of bandwidth while optimizing data transmission time.

GS primarily used a binary encoding strategy with GA to optimize the number of data movements, amount of data movement, and data transmission time in cloud environments. It ignored private datasets and placed all datasets in a cloud datacenter. To compare with GA-DPSO, we modified GS as follows. The fixed storage of private datasets was considered with binary encoding. Moreover, GS considered the bandwidth factor not only in the map from the encoded chromosome to the data placement, but also in the calculation of the fitness function.

Finally, to observe the effect of the preprocessing in section IV, the NGA-DPSO algorithm without preprocessing is used as another comparison algorithm.

### C. Experimental results and analysis

GS, GA-DPSO, and NGA-DPSO belong to the meta-heuristic algorithms. Therefore, they terminate if they maintain their original value after 80 iterations in our experiments. Because the data placement results with the same meta-heuristic algorithm may be different in each experiment, the data transmission time is measured as the average of 100 repeated experiments. The unit of data transmission time is seconds (s), and the experimental results for data transmission time is reduced by 10 times.

Figure 6 shows the data transmission time of different data placement strategies for three kinds of scientific workflows under a basic experiment. In general, GA-DPSO and NGA-DPSO have the best performance. GS is worse compared with GA-DPSO and NGA-DPSO, and the overall performance of DCO-k-means is the worst. Due to the data dependency degree of DCO-k-means being defined based on the pre-placement bandwidth (but not the final bandwidth), there is a gap between the actual data placement and preconceived one. The search scope of GS is relatively limited during each iteration, and it does not adaptively adjust according to the performance of the current chromosome, which results in a worse result compared with GA-DPSO or NGA-DPSO. For Epigenomics and Montage, NGA-DPSO is slightly better than GA-DPSO, and the average data transmission time is reduced by approximately 1.5%. This is mainly due to the fact that the preprocessing affects the final

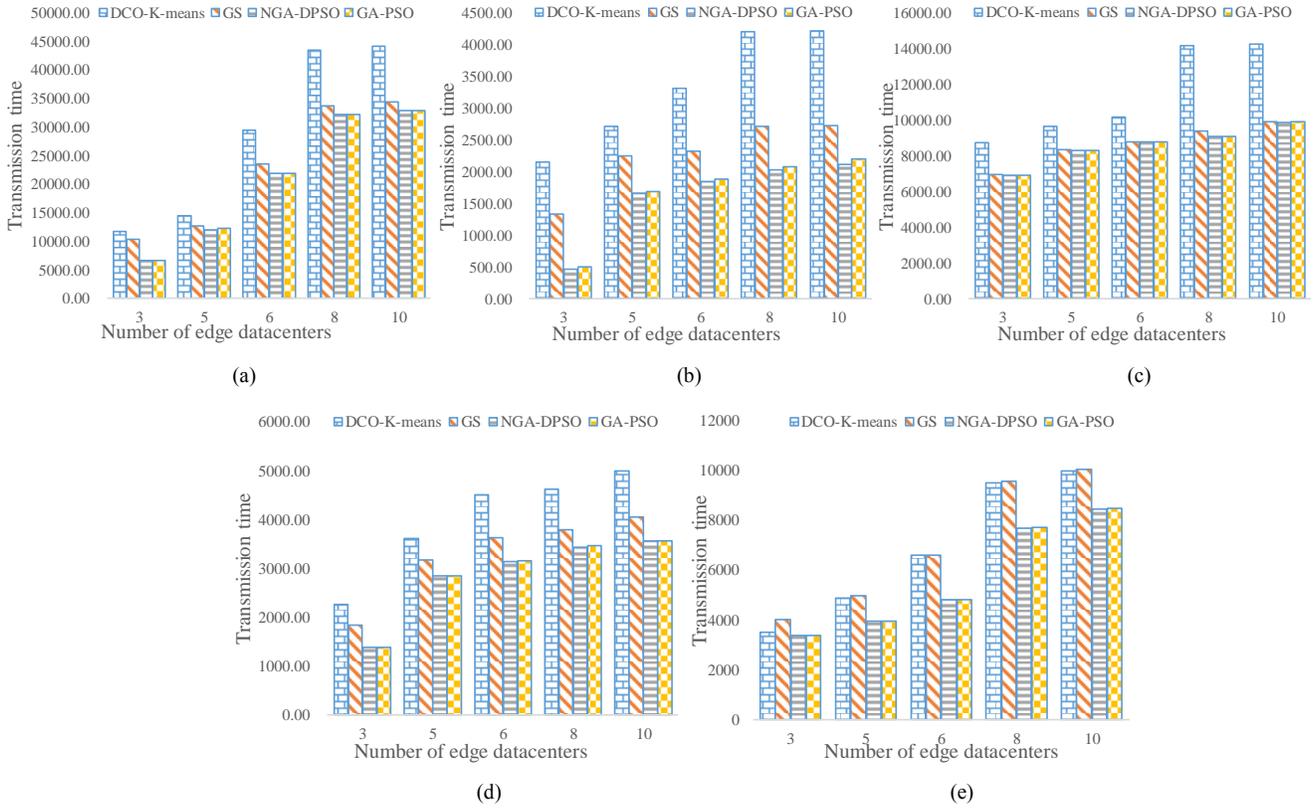

Fig. 7 Data transmission time of different strategies for medium workflows with different numbers of edge datacenters: (a) CyberShake; (b) Epigenomics; (c) LIGO; (d) Montage; (e) SIPHT

data placement result (**Property 1**). The compressed datasets become larger, which may no longer be stored in the original edge datacenter and must be stored in another datacenter with a larger storage capacity. The preprocessing eventually leads to a slight difference between GA-DPSO and NGA-DPSO.

Figure 6(c) shows the data transmission time of different strategies for large scientific workflows under the basic experiment. The strategies in Figure 6(c) cost more data transmission time compared with those in Figures 6(a) and 6(b). This is mainly because of the increase in the number and total amount of workflow datasets, which results in more data transmission across different datacenters. For example, the number of datasets in the small, medium, and large scientific workflow of LIGO is 47, 77, and 1501, and the total size of datasets is 2.47 TB, 4.08 TB, and 82.21 TB, respectively. It costs more time to transmit more and larger datasets in large workflows with the same bandwidth across different datacenters.

Tables 1 and 2 show the average number of iterations and average execution time for the three meta-heuristic algorithms when achieving the optimal result for medium scientific workflows. The average execution time is measured in milliseconds (ms). The average number of iterations of GA-DPSO outperforms NGA-DPSO for Epigenomics and Montage, whose number of iterations can be reduced by approximately 10%. This is mainly due to the preprocessing. The number of datasets of Epigenomics is compressed from 77 to 50, whose compression rate exceeds 35%. Through preprocessing, the number of datasets can be reduced and the encoding space for each particle can be reduced accordingly. Therefore, the number of iterations for searching the optimal result can be significantly reduced. With the compression of the encoding space for each particle, the execution efficiency of GA-DPSO is improved, and the execution time of GA-DPSO is reduced accordingly. From Table 2, it can be seen that the execution time of GA-DPSO is significantly superior to NGA-DPSO for the scientific workflows with high compression ratios, which also benefits from the preprocessing. Regarding the number of iterations and the execution time, GS has the worst performance compared with the other two meta-heuristic algorithms, which is mainly due to the encoding space of GS not being effectively compressed. The search scope of GS is relatively limited during each iteration and does not adaptively adjust itself according to the performance of the current chromosome.

TABLE I
THE AVERAGE NUMBER OF ITERATION WHEN ACHIEVING GBEST FOR THE MEDIUM WORKFLOWS

| Algorithms | CyberShake | Epigenomics | LIGO | Montage | SIPHT |
|---|---|---|---|---|---|
| GS | 482 | 534 | 374 | 472 | 664 |
| NGA-DPSO | 273 | 219 | 273 | 245 | 484 |
| GA-DPSO | 271 | 184 | 275 | 234 | 479 |

TABLE II
THE AVERAGE EXECUTION TIME WHEN ACHIEVING GBEST FOR THE MEDIUM WORKFLOWS (MS).

| Algorithms | CyberShake | Epigenomics | LIGO | Montage | SIPHT |
|---|---|---|---|---|---|
| GS | 89847 | 97902 | 185388 | 105246 | 1051482 |
| NGA-DPSO | 49943 | 76804 | 145821 | 85436 | 853540 |
| GA-DPSO | 50435 | 56851 | 152956 | 75499 | 850142 |

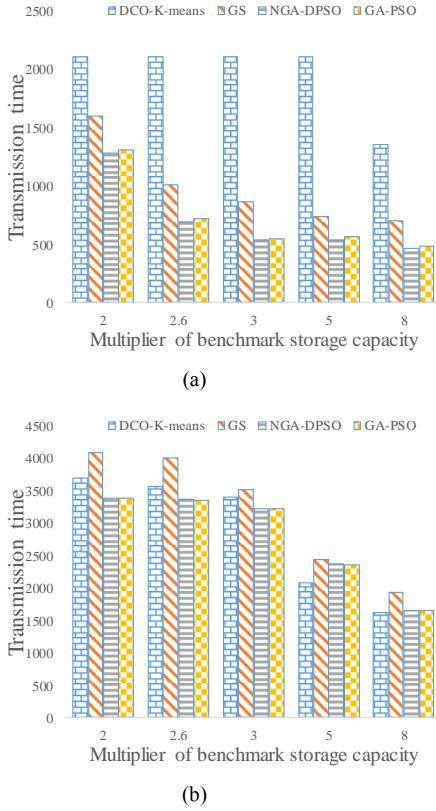

(a)

(b)

Fig. 8 Data transmission time of different strategies for Epigenomics and SIPHT with different storage capacities: (a) Epigenomics; (b) SIPHT

For the number of iterations and execution time of different data placement strategies for small and large scientific workflows, the overall trends are similar to those for medium science workflows. Therefore, the follow-up experiments evaluate the performance of different strategies for only the medium scientific workflows.

To observe the influence of the number of edge datacenters on the performance of different strategies, we adjusted the number of edge datacenters based on the basic experiment. The number of edge datacenters was set to {3, 5, 6, 8, 10}, and the bandwidth between the new edge datacenters was set to 120 M/s. The bandwidth between edge and cloud datacenters was set to 20 M/s. As the number of edge datacenters $|DC|$ increases, the benchmark storage capacity $cap_{benchmark}$ will decrease.

Figure 7 shows the data transmission time of different strategies for medium scientific workflows with different numbers of edge datacenters. As the number of edge datacenters increases, the data transmission time of all data placement strategies increases. The total storage capacity of all edge datacenters remains the same. As the number of edge datacenters increases, the storage capacity of each edge datacenter decreases. As a result, the number or the size of datasets stored in a edge datacenter decreases. The data transmission time across different datacenters increases accordingly. Based on Figure 7, we find that NGA-DPSO and GA-DPSO have the best performance, and DCO-k-means has the worst performance. This is because DCO-k-means divides datasets into different datacenters based on the clustering algorithm, which results in some large datasets being unable to be placed in a suitable datacenter.

The performance of NGA-DPSO and GA-DPSO is almost identical in Figures 7(a), 7(c), and 7(e). This is mainly because the preprocessing has no effect on such workflows (Cyber-Shake, LIGO, and SIPHT) and there are no compressed datasets. However, there is a large gap between NGA-DPSO and GA-DPSO in Figures 7(b) and 7(e), especially for Epigenomics in Figure 7(b). Preprocessing compresses the number of datasets in a scientific workflow, but may affect the final result of data placement (Property 1). When there are a large number of edge datacenters (8 or 10), the impact of Property 1 increases. Figure 7(e) shows the data transmission time of different strategies for SIPHT. It can be seen that the performance of GS is inferior to DCO-k-means. This is because the size of each dataset in SIPHT is similar, and DCO-k-means can obtain a better dataset partitioning result after the clustering algorithm.

In the follow-up experiments, we selected the representative medium scientific workflows (Epigenomics and SIPHT) as the experimental subjects. To observe the influence of the storage capacity of each edge dataenter on the performance of different strategies, we adjusted the storage capacity of each edge datacenter based on the basic experiment. The multiplier of the benchmark storage capacity for each edge datacenter was set to {2, 2.6, 3, 5, 8}.

Figure 8 shows the data transmission time of different data placement strategies for Epigenomics and SIPHT with different storage capacities of each edge datacenter. This hybrid environment includes one cloud datacenter and three edge datacenters. With the increase in the storage capacity of each edge datacenter, more datasets can be stored. The bandwidth between edge datacenters is relatively large, which decreases the data transmission time.

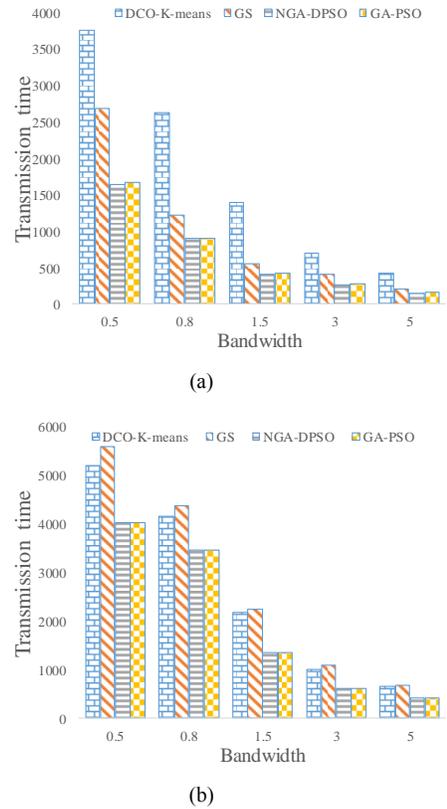

(a)

(b)

Fig. 9 Data transmission time of different strategies for Epigenomics and SIPHT with different bandwidths: (a) Epigenomics; (b) SIPHT

Figure 8(a) shows the data transmission time of different data placement strategies for Epigenomics with different storage capacities for each edge datacenter. From the experimental results, we know that all datasets of Epigenomics are stored only in the edge datacenters with NGA-DPSO and GA-DPSO when the multiplier of the benchmark storage capacity is more than 3. This means that there is no dataset being transmitted to the cloud datacenter, which greatly decreases the data transmission time. However, all datasets of Epigenomics are stored only in the edge datacenters with DCO-k-means when the multiplier of the benchmark storage capacity is more than 8. This is because DCO-k-means first adopted a clustering algorithm to place the initial datasets, then placed the generated datasets based on the data dependency degree. Using such operations, it is impossible to place two large generated datasets of Epigenomics in the same edge datacenter until the multiplier of the benchmark storage capacity is more than eight. Figure 8(b) shows the data transmission time of different data placement strategies for SIPHT with different storage capacities for each edge datacenter. The performance of DCO-k-means is better than GS. This is because there are more datasets (1049), and the size of each dataset is almost the same in SIPHT, which has little impact on the operations of DCO-k-means.

To observe the influence of bandwidth across different datacenters on the performance of different data placement strategies, we adjusted the bandwidth across different datacenters based on the basic experiment. The bandwidth across different datacenters is {0.5, 0.8, 1.5, 3, 5} times faster than the bandwidth in the basic experiment.

Figure 9 shows the data transmission time of different data placement strategies for Epigenomics and SIPHT with different bandwidths across different datacenters. As the bandwidth across different datacenters increases, the speed of data movement increases and the data transmission time decreases significantly. Experimental results show that the increase in bandwidth does not change the final placement of each strategy.

*D. Industrial applications*

Data transmission time plays a decisive role in the user experience of time-sensitive applications. In the application of augmented reality, video application can be transformed into a simple workflow application. The data placement strategy based on GA-DPSO proposed in this paper, combined with the storage resources of edge computing and cloud computing, can effectively reduce the data transmission time of scientific workflow and improve the user experience of the application of augmented reality.

## VI. CONCLUSION

Based on the serious data transmission delays in data placement for a scientific workflow combining edge computing and cloud computing, a time-driven data placement strategy based on GA-DPSO for a scientific workflow was proposed. The experimental results showed that the data placement strategy based on GA-DPSO effectively reduced the data transmission time during workflow execution combining edge computing and cloud computing. While the total storage capacity of all edge datacenters remained the same, the increase in the number of edge datacenters made the data placement more decentralized, and increased the data transmission time. The increase in storage capacity of each edge datacenter effectively increased the number and size of datasets stored in an edge datacenter. Moreover, all datasets could be stored in an edge datacenter with no data transmission time, if the storage capacity of the edge datacenter was large enough. As the bandwidth across different datacenters increased, the data transmission time decreased. However, the final placement for each scientific workflow did not change with different strategies.

In the future, the impact of the proportion of private datasets for a workflow and the impact of each edge datacenter with different storage capacities on the data placement strategies will be considered. In addition, it costs not only time, but also money to transmit data among edge datacenters and cloud datacenters. Therefore, we will comprehensively optimize the data transmission cost of the data placement for a scientific workflow combining edge computing and cloud computing.